# The Role of Conceptual Problem Solving in Learning Physics: A Study in a General Relativity University Course

M. Tuveri[1,2]*, A. P. Sanna[1,2,3], M. Cadoni[1,2]

[1] University of Cagliari, Physics Department, Cittadella Universitaria di Monserrato, S.P. Monserrato-Sestu Km 0,700 - 09042 Monserrato (CA), Italy
[2] Istituto Nazionale di Fisica Nucleare, Cagliari Division, Cittadella Universitaria di Monserrato, S.P. Monserrato-Sestu Km 0,700 - 09042 Monserrato (CA), Italy
[3] Istituto Nazionale di Fisica Nucleare, Roma Division, Piazzale Aldo Moro 5, 00185, Roma, Italy

*Corresponding author: matteo.tuveri@ca.infn.it

**Abstract**

Effective physics learning, especially in complex topics, requires balancing mathematical formalism with conceptual understanding. Conceptual problem-solving involves connecting math to physical reality, and using an epistemological framework like problem framing helps students justify their mathematical decisions. This approach deepens students' understanding by linking theory to practice and enhancing their reasoning skills. This study explores the effectiveness of conceptual problem-solving in learning complex topics like general relativity (GR) through a pedagogical framework that emphasizes the integration of qualitative and quantitative reasoning. We present a case study conducted at the University of Cagliari in 2021 and 2022, examining how students construct problem frames and how this influences their conceptual understanding of GR. Findings indicate that students who effectively integrate conceptual reasoning with mathematical formalism demonstrate a deeper grasp of physical principles and enhanced problem-solving capabilities. The research underscores the importance of symbol sense and the iterative nature of problem framing, suggesting that an integrated approach—combining visual, symbolic, and natural language representations—can improve students' conceptual engagement. Furthermore, the methodology offers instructors valuable insights into students' thinking processes, supporting more effective and targeted feedback.

## 1. Introduction

The effective teaching of physics, particularly in complex areas such as general relativity (GR), requires a careful balance between mathematical formalism and conceptual understanding [1]. Traditional problem-solving in physics often focuses heavily on mathematical computations, yet the underlying conceptual knowledge—how principles and ideas shape the structure of a solution—is equally critical [2,3]. In recent years, research has shown that focusing on conceptual problem solving can improve students' understanding of physics and their ability to apply knowledge in novel contexts [4-6]. Indeed, a key aspect of expert problem solving in physics is the development of symbol sense, which refers to the ability to reason conceptually about symbols and their relationships within equations [7]. Symbol sense is not just about manipulating symbols. It also involves interpreting the conceptual meaning behind the symbolic relationships, generating expressions based on intuitive and conceptual understanding, and knowing when and how best to use this understanding in problem-solving contexts [8-11]. For example, in general relativity (GR), understanding how the symbols in Einstein's field equations relate to physical phenomena is not just a matter of mathematical computation but of interpreting the meaning behind those symbols in a physical context.

In this regard, blended processing—the integration of conceptual reasoning with mathematical formalism—can offer a more productive approach to problem solving than simply using equations as computational tools [12-14]. By blending conceptual and mathematical thinking, students are able to develop a deeper understanding of the physical systems they are modeling and are better equipped to reason about them both qualitatively and quantitatively [15]. This approach aligns with the idea that mathematical equations should not be seen merely as computational procedures but as representations that also carry deep conceptual meaning [16-18].

One framework that supports this view is the concept of symbolic forms which has been proposed by Sherin (2006) [19]. According to the author, symbolic forms connect mathematical equations to intuitive conceptual ideas. A symbol template represents the general structure of an equation—its symbols and operations—without specifying the actual values or variables. This template is blended with a conceptual scheme, which refers to the intuitive, often informal ideas that underlie the mathematical structure. These conceptual schemas are typically drawn from every day, non-academic knowledge. For instance, the idea that "a whole consists of many parts" could be a conceptual schema that helps students understand systems with multiple interacting components, such as gravitational fields in GR.

This blending of symbol templates and conceptual schemas results in a unified way of thinking, where reasoning is neither purely formal and mathematical nor purely conceptual but a blend of both [10]. This integrated approach allows students to apply their intuitive understanding of physical phenomena while simultaneously engaging with the mathematical structure that describes them [20,21]. As such, the use of symbolic forms in problem solving provides a more holistic understanding of physical systems, fostering deeper comprehension and greater flexibility in tackling complex problems [22]. Moreover, the ability to move fluidly between conceptual reasoning and mathematical formalism is a hallmark of expertise in physics [12,13,23]. Research has shown that students who are able to leverage both aspects of reasoning—conceptual and mathematical—are more adept at translating mathematical solutions into physical understanding [24-27]. This skill is particularly important in advanced topics like GR, where the complexity of the equations requires not only mathematical rigor but also a conceptual grasp of the physical principles involved.

Conceptual problem-solving in physics requires more than just applying mathematical formulas; it involves understanding how math relates to the physical world [28]. Indeed, as emphasized by Paul Hewitt in his long-standing call in 1982 Millikan Lecture [29], physics instruction should focus on conceptual aspects rather than rote on quantitative problem-solving and algorithmic manipulation at the expense of students' actual meaning-making of physical phenomena. Hewitt pointed out that while students may learn to plug numbers into equations, they often fail to develop a robust conceptual framework - "an inner coherence of knowledge" - that allows them to interpret and reason about physical situations beyond routine computations. He further emphasized that physics instruction should aim to build *conceptual fluency* (students' ability to reason qualitatively about what is going on) and *representational flexibility* (the ability to move between different representations – verbal, mathematical, diagrammatic) rather than simply producing correct numerical results. In doing so, Hewitt anticipated many of the later developments in physics-education research that focus on students' epistemological and representational frames: instructors are encouraged to create learning environments in which students attend to *what kind of thinking* they are doing, *how* they are reasoning, and *why* – not just *what* they compute.

Rephrased in modern context, the implementation of a strong epistemological framework could help students in this process, giving them a methodology to justify their math decisions and connect abstract concepts to real-world phenomena [30]. In this paper, we use the term epistemological framework to refer to the system of assumptions, values, and criteria that guide how knowledge is constructed and justified within a discipline. In contrast, epistemic framing describes how individuals or groups momentarily perceive "what kind of activity" they are engaged in—whether they are doing conceptual reasoning, applying formulas, or interpreting representations [12,13]. Epistemic framing is therefore a dynamic process that reflects students' shifting orientations toward knowledge and problem solving. Understanding these shifts can help physics instructors design learning environments that support deeper conceptual engagement and flexibility across different modes of reasoning.

When applied to problem solving, this framework is better known as problem framing. It shapes how students approach problems and perceive their mathematical methods [12-18]. This approach enhances students' ability to reason and connect theory with practice [30]. By focusing on how students frame their problem-solving, educators can deepen their understanding of both mathematical processes and physical concepts.

This paper presents a pedagogical framework aimed at enhancing conceptual problem-solving skills in GR for students in a master's-level physics course. By emphasizing the role of qualitative reasoning alongside quantitative techniques, we explore how this approach can support students' ability to construct coherent explanations and navigate complex problems in GR. Specifically, we aim to assess how students use conceptual knowledge to select appropriate principles, justify their choices, and transition from qualitative reasoning to quantitative problem solving. Through the use of specific solving strategy which points towards developing students' epistemological framing and expert-like way to face with problems, this research examines the ways students blend conceptual understanding with the formalism required for solving GR problems. We were interested in investigating the following aspects (research questions):

RQ1: how do students structure their problem framing within conceptual problem solving?

RQ2: how does conceptual problem framing help students in conceptual learning of GR?

RQ3: is our methodology efficient in investigating students' conceptual problem solving and framing competences?

In the following sections, we will explore the pedagogical goals that inform this approach, review the theoretical foundations of conceptual problem solving in physics education, and describe the methodology used to evaluate students' understanding in the context of GR. The results are then discussed in terms of their implications for teaching and learning in the discipline of physics, with a focus on conceptual and symbolic reasoning. Additionally, this paper addresses the potential benefits for instructors in gaining insight into students' thought processes and developing targeted instructional strategies.

## 2. Theoretical framework

To foster a deeper understanding of physics, it is essential for students to develop the ability to think like experts in physics, which requires more than just solving equations [25]. Van Heuvelen (1991) suggested that students can achieve this by being given opportunities to reason qualitatively, engaging in "translations" between verbal, pictorial, and physics representations before transitioning to the mathematical form of the problem [23]. This process encourages students to develop a mental model of the problem, grounded in intuitive understanding, before diving into the quantitative aspects. By reasoning qualitatively first, students can build a stronger conceptual foundation that enhances their ability to apply mathematical formalism later in the problem-solving process [26].

Building on this approach, Redish (1994) emphasized the importance of integrating constructivist learning theories with the practice of teaching physics [17]. According to Redish, students should not only engage in qualitative reasoning but also have the opportunity to construct and refine mental models that represent the underlying physical concepts. As students develop these models, they are better equipped to apply them to solve complex problems. Redish also highlighted the importance of conceptual change—a shift in students' understanding facilitated through active engagement and discussions. When students work together and resolve conflicts in their reasoning, they strengthen and refine their mental models, allowing for more sophisticated problem-solving strategies. Indeed, effective problem-solving in physics requires students to understand the connection between mathematics and physical reality, going beyond simply applying formulae.

Developing a strong epistemological framework helps students justify their mathematical choices and interpret abstract concepts in the context of real-world phenomena. This involves problem framing, where students adopt mental frameworks to approach problems [12,13]. These frames are dynamic, and students shift between them depending on the problem and errors encountered. Understanding problem framing is crucial for addressing difficulties and improving problem-solving strategies [30-32]. Indeed, in problem-solving, students should shift between different epistemological frames based on their focus and understanding of the task [33].

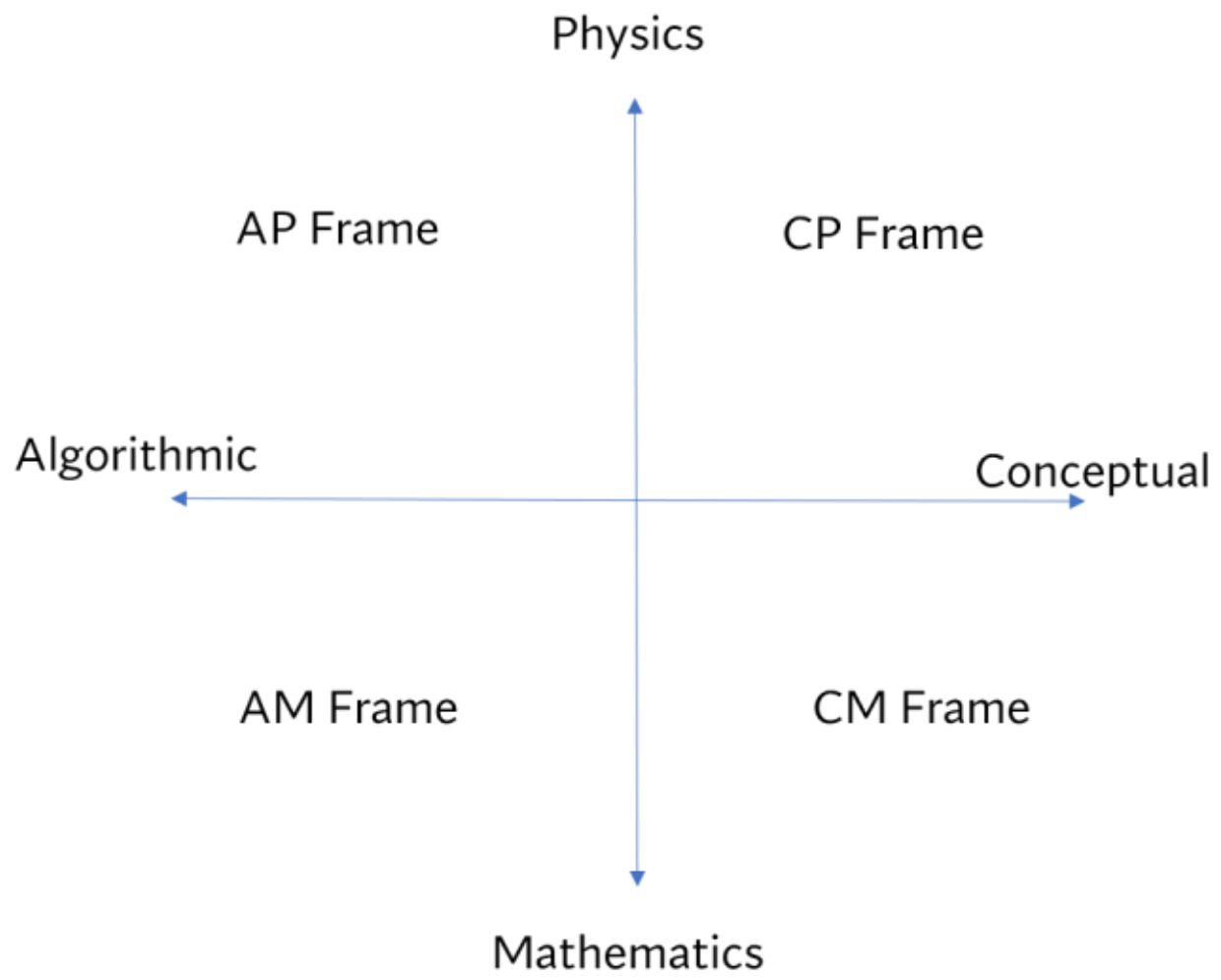

Fig. 1 Problem framing as described by Nguyen et al. [30]. The two axes divide the space into four frames, namely Conceptual Physics (CP), Algorithmic Physics (AP), Algorithmic Math (AM), Conceptual Math (CM).

For this reason, as emphasized by Nguyen et al. [31], an epistemological framing which could also take into account cooperative learning pedagogies should include a deep interplay between conceptual and algorithmic frames. The authors developed a model consisting on four different frames, see Fig. 1: Conceptual Physics (CP), where students discuss physical concepts and plan their solutions; Algorithmic Physics (AP), which involves translating physical understanding into mathematical formulations; Algorithmic Math (AM), where students perform mathematical procedures to obtain results; and Conceptual Math (CM), which emphasizes applying mathematical rules without delving deeply into the detailed problem. These frames evolve as students interact in group settings, revealing knowledge gaps and leading to more productive problem-solving. Group discussions and negotiations help students refine their frames, fostering deeper understanding and improving problem-solving efficiency. Through this process, students not only solve problems but also develop stronger conceptual frameworks. Understanding these shifting frames can guide educators in creating more effective learning environments.

Building on recent work by Kapodistrias et al. (2025), we can describe students' shifts between epistemic frames as instances of purposeful transformation – intentional reorganizations of reasoning that serve a specific disciplinary purpose [34]. In their study of mathematical reasoning in cosmology, Kapodistrias and Airey identify three key characteristics of such transformations:

(1) a narrowing of interpretive possibilities, through which the range of meanings associated with a representation or equation becomes more focused and useful for reasoning;

(2) a foreground–background movement, whereby some aspects of a situation are highlighted while others are temporarily set aside to make disciplinary relationships more visible; and

(3) a purposeful direction, indicating that the transformation is guided by the questions, goals, and values of the discipline itself.

In this sense, the transitions observed between the Conceptual Physics (CP), Algorithmic Physics (AP), Algorithmic Math (AM), and Conceptual Math (CM) frames can be understood as purposeful transformations in students' disciplinary thinking. Each transition reflects a deliberate reorganization of how knowledge is structured and applied to achieve conceptual coherence. Incorporating this perspective helps to interpret framing dynamics not only as cognitive shifts but also as goal-oriented restructurings of reasoning that bring students' approaches closer to expert disciplinary practice.

The value of this approach is supported by instructional strategies that prioritize qualitative analysis and multiple representations of physical problems [35]. Research by Heller and Reif (1984) found that explicitly teaching students problem-solving strategies that incorporate qualitative reasoning and multiple representations led to improved problem-solving skills [36-39]. This aligns with the idea that students should be encouraged to describe the physical phenomena qualitatively, identify underlying principles, and develop strategies before moving on to the formal mathematical solution. Moreover, visual representations are also fundamental [40]. Indeed, an appropriate diagram is a required element of a solution building process in physics problem solving and it can transform a given problem into a representation that is easier to exploit for solving the problem. Additionally, strategies like qualitative strategy writing—where students describe how they would approach solving a problem—have been shown to enhance both problem-solving skills and conceptual understanding [4,41-46]. This task allows students to articulate their reasoning in qualitative terms, focusing on what principles apply, why they are relevant, and how they can be used to solve the problem. It is an essential part of developing expert-like thinking, as students learn to connect their intuitive conceptual understanding with the formal mathematical structures that characterize physics problems [47,48].

A further extension of these strategies is found in modeling instruction, which requires students to engage in discussions, resolve conflicting ideas, and refine their models of physical phenomena [49,50]. Through this iterative process of discussion and model adjustment, students enhance their understanding of the principles they are learning and learn to apply them more effectively in problem-solving contexts. This process of discussion and model-building creates a more dynamic learning environment that promotes conceptual change and facilitates deeper understanding.

The importance of familiarity with problems in promoting conceptual understanding is also crucial [51-53]. Learning through familiarity, according to Alant, allows students to connect new problems with prior knowledge, making it easier to grasp the underlying physics principles. When students are exposed to closely related problems, they develop a more coherent understanding of the concepts and are better prepared to solve more complex challenges. This approach fosters the development of robust mental models that can be applied across various contexts, reinforcing the connection between intuition and formal reasoning.

A useful framework for integrating these ideas into physics education is Greeno's "extended semantic model" [52]. This model highlights four key domains that students must navigate to develop expertise in conceptual problem-solving. First, there's the concrete domain. This is all about the physical objects and events students encounter in the real world. It's the stuff they can touch, see, or experience directly. Then, as students move forward, they start working with the model domain. This is where they deal with representations of real-world phenomena—models or abstractions that help explain how things work in a simplified way. As they deeply grapple physics, they enter the abstract domain. This is where they engage with concepts, laws, and principles—the fundamental scientific reasoning that underpins everything. Finally, there's the symbolic domain, which is basically the language of physics. This is where math comes into play: algebra, equations, and all the symbols that students use to describe relationships and solve problems. By engaging with each of these domains, students can connect their qualitative understanding of physical phenomena with the abstract concepts and symbolic representations that are central to physics. This approach encourages students to develop a more holistic understanding of the subject, allowing them to reason both conceptually and mathematically when solving problems.

Finally, let us remark that the pedagogical and cognitive dimensions of teaching problem-solving have long been addressed in both science education and physics education research [53-55].

This study builds on this foundation by exploring how students shift between frames during cooperative problem solving.

## 3. Design of the activity

Building upon prior research in the field [41-46,50] and following the conceptual framework outlined above, we developed a conceptual problem-solving activity aimed at engaging students in learning GR. To create a more relatable learning experience, we selected contextualized problems that linked physics concepts to real-world phenomena. This approach encourages students to immerse themselves in practical scenarios, prompting them to question which physical principles to apply and why, and what specific concept should be used to solve the problem [42-46].

The problems were based on four conceptual topics in GR, grouped under two main macro-themes, see Table 1: (1) the geometric nature of gravity and the properties of spacetime, and (2) gravitational phenomena as described by GR. The activity had two primary objectives: to provide students with a relatively comprehensive and in-depth understanding of key issues in GR, and to strengthen and broaden their knowledge of physics and mathematics through engaging and motivating topics that connect mathematical formalism with physically and contextually meaningful phenomena. Based on these themes, we selected four core conceptual issues in GR—two for each macro-theme:

- the propagation of signals in flat and curved spacetime (e.g., the Doppler effect in special and general relativity);
- the properties of curved geometries and related phenomena (e.g., geodesic motion and the parallel transport of a vector);
- phenomenology of curved spacetime, with attention to contrasts between flat and curved geometries (e.g., light deflection);
- gravitational redshift caused by the gravitational field of a black hole.

Table 1. Description of meetings, topics, and learning goals of conceptual problems given to students during the training activity.

| Meeting | Topic | Learning goals |
|---|---|---|
| 1) Introduction to conceptual problem solving and how to face with problems. Solving activity: collegial | Doppler effect in Special and General Relativity, GPS | Interplay between mathematics and physical conceptual understanding |
| 2) Solving activity at home: group working or alone | Parallel transport of a vector, Geodesic motion | Equivalence principle and the physical meaning of Einstein's Equations |
| 3) Solving activity at home: group working or alone | Geodesic deviation, light propagation in curved spacetime | GR astrophysical scenarios: black holes |
| 4) Solving activity at home: group working or alone | Gravitational redshift | GR-everyday life applications: gravitational redshift |

To situate our design with respect to the Nguyen et al. framework [31], we contextualized problem framing within the problem-solving activities routinely used in the Physics course at the University of Cagliari. From the first-year introductory course in mechanics and thermodynamics onward, students are exposed to problem-solving and context-enriched problems [41–45]. In this tradition, instruction typically follows the methodology introduced by Heller et al. [47,48], namely a five-step iterative procedure: focus the problem; describe the problem; plan a solution; execute the plan; evaluate the solution. We incorporated the frames in [31] by embedding them into the aforementioned five-step procedure. We prepared a template (see Appendix A) to present our methodology and as the instructional vehicle through which students were introduced to framing and framing shifts. While we did not name this explicitly as "problem framing" nor require memorization of the taxonomy, the template was designed to (i) guide students' reasoning across CP/AP/AM/CM during problem solving and (ii) make prospective framing shifts traceable in their written artefacts for subsequent analysis; the taxonomy itself was not an assessment target.

The idea is that, rather than diving straight into equations, students should be guided to begin with qualitative reasoning, in line with expert practice. The process opened by visualizing the physical situation (*step 1: drawing a diagram to clarify the physical setting*), thereby creating a bridge from an initial conceptual account to later formalization (*Conceptual Physics*). Students then described the phenomena, justified the relevant principles, and articulated how these principles could be used (*step 2 – Algorithmic Physics*). Emphasizing conceptual physics at the outset aligned the activity with the epistemological framing framework [31] and encouraged approaches commonly observed in expert physicists.

Once a qualitative grasp was established, students transitioned to the quantitative side, moving from conceptual to algorithmic frames. They planned a solution strategy (*step 3; Algorithmic Physics and Mathematics*) by identifying which models or equations would be needed (drawing as appropriate on Conceptual/Algorithmic Math), executed the solution (*step 4; Algorithmic Mathematics*), and finally evaluated the result (*step 5; Conceptual Mathematics and Conceptual Physics*), checking coherence with the earlier conceptual account and with the identified principles. This five-step structure, grounded in prior research on problem solving [36,41–43], was intended to maintain consistency between symbolic expressions and the physical world, ensuring that formal manipulation served the underlying physical reasoning.

In addition to the main problem-solving task, we introduced a problem-categorization exercise. Students were given a contextualized problem without being required to solve it; they selected the most relevant principles from a provided list, focusing attention on conceptual discernment without the pressure of a full solution.

The same template (Appendix A) hosted all tasks, organizing students' written reasoning and supporting self-monitoring of shifts between conceptual and algorithmic reasoning in physics and mathematics, while keeping the focus on the course's disciplinary content and methods.

## 4. Methodology

### *4.1 Implementation of the research activity*

The research activity was conducted in 2021 and 2022 at the Department of Physics of the University of Cagliari (UniCa), Italy, as part of the GR course for the Master's degree in Physics. The course was taught by one of the authors (the lecturer) in both years, and the syllabus and methodology were identical across the two cohorts.

The GR course has been part of the theoretical curriculum in "Physics of Fundamental Interactions" of the Master's program in Physics at UniCa since 2018. Students in other tracks (e.g., astrophysics, particle physics, cosmology) may choose it as an elective. The course is generally attended by students from both theoretical and experimental physics as well as mathematics undergraduate programs.

Course materials included selected textbooks [56–58] and lecture notes. Topics ranged from differential geometry and the principles of equivalence and general covariance, to Einstein's equations and GR phenomenology in relativistic astrophysics, such as gravitational lensing, black holes, gravitational waves, and cosmology.

The course consists of 48 contact hours, delivered over 24 sessions. Teaching is divided between lectures and training sessions. The latter consist of eight two-hour meetings held during regular class time. Until 2021, these training sessions relied on traditional textbook exercises, with no implementation of conceptual problem-solving methodologies. These standard exercises aimed to support students in acquiring core competencies in GR. Outside the research activity, students were typically exposed to lecture-based instruction, board derivations, and end-of-chapter problems, with limited emphasis on diagrammatic or conceptual representations. These activities did not include an explicit, methodological guide in conceptual problem-solving methodology and did not require students to mark or reflect on frame shifts in their written work. This contextual difference is relevant for interpreting students' comments about the novelty of the intervention.

Our educational program, designed as part of a research initiative, was introduced as an experimental teaching activity in 2021 and 2022. During these years, the tutoring sessions were led by one of the authors—at the time a PhD student specializing in GR—who had been selected by the course lecturer. The research component was discontinued after 2022 due to the unavailability of further PhD support. Notably, the same tutor led the activities in both years, ensuring consistency in the topics, methodologies, and implementation.

The study was embedded in four training sessions. The first session introduced the methodology and the conceptual problem-solving approach, together with the template in Appendix A. Students could work individually or in groups. For each session, the tutor uploaded one exercise to a shared online folder, and students had one week to complete the task. At the end of the week, solutions were presented and discussed in a synchronous meeting. Throughout the week, the materials remained accessible to support asynchronous engagement: students could post comments, request clarifications about the physical scenarios, and share reflections. The full set of problems is listed in Appendix B.

The first meeting was tutor-led (aims of the research program, description and use of the template, expectations concerning students' learning process in both problem-solving methodology and conceptual learning). In the subsequent meetings, the tutor collected templates, facilitated whole-class debriefs during the synchronous activities, and answered questions posted asynchronously, but did not co-solve the tasks or steer students' framing during the week.

Consistent with our theoretical framework and design, instruction emphasized pictorial and diagrammatic representations as a primary entry point to the physics of each problem. Visual representations were highlighted for identifying relevant quantities and for reasoning about both system dynamics and system states. This visual–conceptual stance was developed collaboratively within the research team and was introduced to students during the first session, along with an explanation of the proposed methodology.

On the same day, students were trained to use the problem-solving template (Appendix A) and were asked to apply the five-step procedure. The aim was to emphasize the interplay among frames when tackling problems. The tutor illustrated the approach with a fully worked example. All problems were authored by the tutor and reviewed by all co-authors. During whole-class problem solving, the tutor took field notes documenting how the template should be used to mark framing moves and to identify when shifts across frames naturally emerged in students' unfolding reasoning.

By the end of the training activity, students were expected to:

1. Produce a diagrammatic/pictorial sketch of the physical situation and identify the relevant physical quantities (Step 1; CP);
2. Write a qualitative account justifying which physical principles apply and why (Step 2; CP→AP);
3. Outline a solution plan indicating when to pass from qualitative to algorithmic reasoning and which relations/models to use (Step 3; AP→AM);
4. Translate the plan into symbolic form and carry out only the necessary mathematical steps (Step 4; AM);
5. Evaluate the result for coherence with the qualitative account and for basic mathematical soundness (e.g., units, limits, functional dependencies) (Step 5; CM & CP).

The tutor modeled these expectations on a worked example and asked students to mark frame shifts within the template (e.g., through headings or brief notes), to separate qualitative justification from algebraic execution, and to start from a diagram before formal manipulation.

Finally, students' performance in the problem-solving activity served as a formative assessment and qualitatively contributed to their final exam grade. The final exam consisted of a presentation (with slides) on a selected topic, followed by an oral discussion with the lecturer. The tutor qualitatively assessed each student's work (as positive, negative, or neutral according to the analysis criteria described in Section 4.3) and reported to the lecturer, who used this formative input to confirm or adjust the final grade.

*4.2. Sample*

In 2021, twelve students attended the first two meetings, and ten completed the full training sequence. In 2022, nine students participated in the activity. During 2021, some students chose to work in groups for Exercise 2, forming three groups (two groups of three students and one group of two). In contrast, in 2022, all students worked individually. All participants held a bachelor's degree in physics. Informed consent for the use of student work in research was obtained from all participants. Consent was not required for participation in the learning activities themselves. Data were analyzed anonymously and reported in aggregate form. At the time of the study, ethical approval from the department was not required for this type of educational research.

*4.3 Analysis of the elaborates*

This study adopts a qualitative case study approach, which is well suited for exploring educational processes in depth and within context [59,50]. As outlined by Denzin and Lincoln [61], qualitative inquiry emphasizes the situated nature of knowledge and the interpretive role of the researcher, making it particularly appropriate for analyzing students' written solutions, representational choices, and frame transitions across two implementations (2021 and 2022). The dataset consists of students' problem-solving artifacts, which were examined to trace epistemic and conceptual reasoning in an authentic classroom setting. We collected and analyzed the completed templates from each student at the end of every training session to understand their approach to problem solving.

The primary aim of our study was to evaluate the effectiveness of our instructional design—specifically the use of conceptual problems—in shaping students' problem framing. To this end, we interpreted students' solutions using the epistemic framing framework described in [31] and outlined in Section 2. Additionally, we analyzed students' problem-solving strategies by tracking how they integrated algorithmic, conceptual, and symbolic reasoning. Following [31], we define a frame shift as a transition from one activity to another that reflects a change in epistemic framing. A shift occurs when students begin to approach a task differently than before. In contrast, a transition between two different actions within the same frame does not constitute a shift, as the underlying framing remains unchanged—even if students attempt different strategies.

Unlike in [31], due to the post-pandemic context during which the research was conducted, we were unable to perform video analysis to observe group work or student interaction. As a result, we could not categorize frame shifts based on peer interaction, as done by Nguyen et al. [31]. Furthermore, the tutor did not participate in solving the problems and played no role in influencing students' framing or triggering shifts. The tutor only collected students' templates during synchronous meetings and provided feedback afterward. Accordingly, and in line with [31], we focused exclusively on student-initiated shifts and natural frame transitions that emerged during individual problem solving. In our framework, these shifts are not merely technical changes but signal epistemic transformations, where students reinterpret the problem using new conceptual or operational structures. Monitoring these shifts provides insight into the depth of students' reasoning and helps assess whether they are exhibiting novice-like or expert-like thinking.

To identify these shifts, we examined linguistic markers, formulas, and graphical representations—such as specific phrases, symbolic transitions, or drawings—that revealed movement from one frame to another. The types of frame transitions and the indicators that triggered them, as identified in our study, are as follows:

- Conceptual Physics → Conceptual Physics: A non-transition observed when students' work contains only principles and conceptual reasoning, without numerical solutions or formula use.

- Conceptual Physics → Algorithmic Physics: Typically marked by the introduction of formulas, often following a conceptual or pictorial representation.
- Algorithmic Physics → Conceptual Physics: Occurs when students, after applying a formula, invoke physical principles (e.g., equivalence principle, general covariance) or use drawings and diagrams to continue the reasoning.
- Algorithmic Physics → Algorithmic Math: Identified when physical expressions are reformulated mathematically (e.g., using inverse functions or simplifying equations).
- Conceptual Math → Algorithmic Physics: Happens when mathematical concepts (e.g., circle geometry) are used to define or calculate physical quantities.
- Algorithmic Physics → Conceptual Math: Detected when formulas are followed by reflection on mathematical relationships, such as properties of tensors.
- Conceptual Math → Conceptual Physics: Arises when mathematical ideas are recontextualized within physical concepts (e.g., from geodesic to motion).

*4.4 Assessment of solutions*

To assess the quality of students' solutions, we considered a problem to be satisfactorily solved when the response was complete, coherent, and both conceptually and formally accurate, addressing all aspects of the task rigorously. The assessment criteria are based on the rubric developed by Heller et al. [55], which identifies six stages of expert problem solving in physics:

1. Conceptual Understanding: Demonstrating correct interpretation of physical concepts and their interrelations.
2. Completeness of Representation: Including all essential elements, such as diagrams.
3. Consistency Between Representations and Formulas: Ensuring equations align with the physical model and representations.
4. Solution Planning: Outlining a clear strategy before applying algebraic operations.
5. Logical Progression: Presenting a well-structured reasoning sequence, with numerical substitution only after isolating the unknown.
6. Mathematical Correctness: Using algebraic manipulations that are valid, consistent, and free of structural errors.

For clarity, we operationalized "expert-like performance" solely through the six Heller et al. categories mentioned above [55]. No frame counts of shifting were considered at this stage. Moreover, we classified a solution as *successful* if it met *all* six criteria, with conceptual and formal coherence across the complete chain of reasoning; partially successful if one or more criteria were weak or missing but the main line of reasoning was recoverable; and *unsuccessful* if conceptual understanding and/or consistency were insufficient to support a coherent solution.

The classification also supported the qualitative analysis of students' performance and informed the lecturer's formative evaluation (Section 4.3). An overall judgment was assigned as follows: positive when solutions were successful across all proposed activities; negative when unsuccessful outcomes outnumbered successful ones; and neutral when the numbers of successful and unsuccessful outcomes were comparable.

Crucially, this overall classification was assigned prior to and independently from any analysis of frame activation or transitions. Only after coding solution quality did we examine frame usage to characterize reasoning dynamics.

To align our framing analysis with these established criteria, we interpreted the six phases above as actions occurring within specific frames. This allowed us to construct an integrated methodology linking traditional problem-solving analysis to the epistemic framing framework. Specifically (see also Fig. 2):

- Conceptual Understanding and Completeness of Representation were mapped to the Physical Conceptual Frame, which involves discussing and mapping physical concepts and representations to plan a solution.

- Consistency Between Representations and Formulas, Completeness of Representations, and Solution Planning were associated with the Physical Algorithmic Frame, where students translate conceptual understanding into an algorithmic solution strategy.
- Solution Planning and Logical Progression also mapped to the Mathematical Algorithmic Frame, representing the phase where students execute mathematical procedures in a structured manner.
- Mathematical Correctness was linked to the Mathematical Conceptual Frame, where students reflect on mathematical properties and relationships.

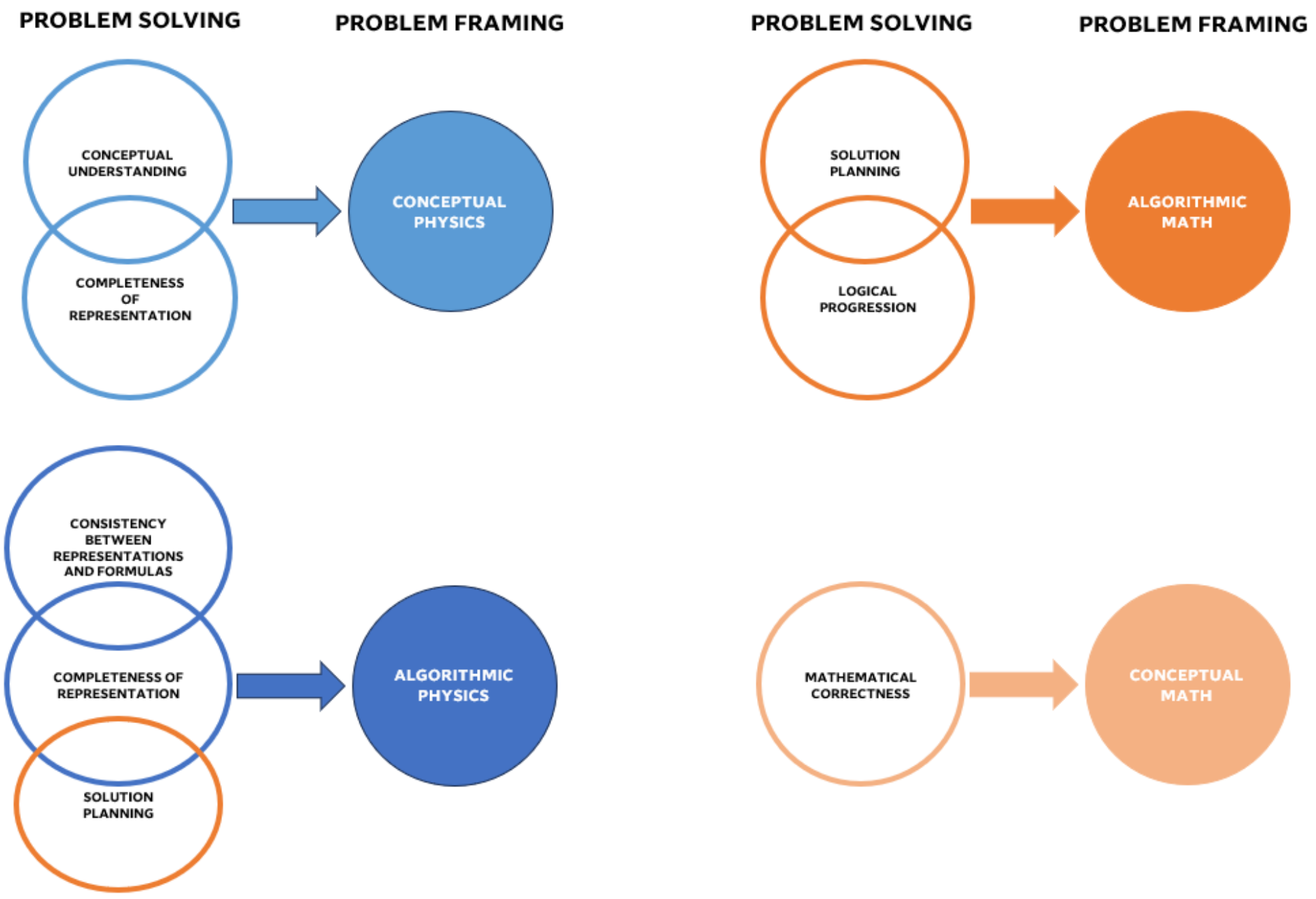

Fig. 2. Mapping of Heller et al.'s [55] six expert problem-solving phases onto the epistemic framing framework proposed by Nguyen et al. [31]. This figure illustrates our proposal to reinterpret the six evaluation steps as foundational actions within specific frames. Notably, solution planning spans both the physical algorithmic and mathematical algorithmic frames, while completeness of representation contributes to both the physical conceptual and physical algorithmic frames.

In our assessment scheme, the *Algorithmic Math* frame designates students' performance in mathematical procedures to obtain results as in [31]. We intend it as the procedural aspects of planning and implement a mathematical solution (*Solution Planning* and *Logical Progression)* in problem-solving. These primarily concern the ordered execution of mathematical steps, by also taking into account technical aspects such as symbolic manipulation, substitution, algebraic isolation of terms. By contrast, we linked *Mathematical Correctness* to *Conceptual Math* frame because in our coding "correctness" denotes structure-level mathematical reasoning (e.g., dimensional/exponent checks, functional dependencies, limits, tensorial properties) that interrogates the meaning of expressions rather than their mere execution. Put differently, the AM frame captures how procedures are carried out. The *Conceptual Math* frame evaluates what the obtained expressions say in mathematical terms and whether that meaning coheres with the intended structure. This delineation aligns with our definitions of AM ("perform procedures to obtain results") and CM ("reflect on mathematical properties and relationships").

By combining the frame analysis with traditional problem-solving evaluation, we reconstructed the cognitive processes that characterize expert reasoning. This combined framework served as an analytical tool for assessing both the completeness and accuracy of students' framing and solutions. Completeness was considered closely linked to correctness. Different levels of completeness were identified: when answers were partial or omitted key steps, solutions were labeled as incomplete, even if partially correct. We also distinguished between solutions containing conceptual errors and those with formal inaccuracies. Once the classification of correctness and completeness was complete, a second round of analysis focused on frame activation during problem solving. This involved a detailed reading of each student's work to identify language (written), formulas, and diagrams that revealed the activation of a specific frame or a transition between frames.

Finally, we collected feedback from students at the end of the course to evaluate how the methodology influenced their learning process. These results are discussed in the following section and summarized in Tables 2–7. Where necessary, data were disaggregated by cohort (2021 and 2022) to highlight differences in student responses and frame use.

## 5. Results and discussion

The analysis revealed significant differences in how students approached problem-solving. Successful strategies were mostly found in the first year 2021. Generally, students exhibited a well-structured approach to framing during the activity, as shown in Tables 2-4.

Table 2. Frequency of shifts in Problem 1 (on the left) and Problem 2 (on the right) from starting frame to following frame. Data refers to 2021 cohort. CP stems for "Conceptual Physics", AP for "Algorithmic Physics", AM for "Algorithmic Mathematics", and CM for "Conceptual Mathematics". Empty cells indicate frames that were absent from the analysis.

| | Following frame | | | | | | | Following frame | | | | | |
|---|---|---|---|---|---|---|---|---|---|---|---|---|---|
| Problem 1 | | CP | AP | AM | CM | Row total | | Problem 2 | CP | AP | AM | CM | Row total |
| Starting frame | CP | | 8 | 2 | 5 | 15 | Starting Frame | CP | 1 | 7 | 7 | 7 | 22 |
| | AP | 1 | 3 | 2 | 2 | 8 | | AP | 9 | | 15 | 6 | 21 |
| | AM | | | | | | | AM | | 1 | 1 | 1 | 3 |
| | CM | 2 | 2 | | 1 | 5 | | CM | 1 | 3 | 3 | 2 | 9 |
| | Column total | 3 | 13 | 4 | 8 | | | Column total | 11 | 11 | 26 | 16 | |

Table 3. Frequency of shifts in Problem 1 (on the left) and Problem 2 (on the right) from starting frame to following frame. Data refers to 2022 cohort. CP stems for "Conceptual Physics", AP for "Algorithmic Physics", AM for "Algorithmic Mathematics", and CM for "Conceptual Mathematics". Empty cells indicate frames that were absent from the analysis.

| | Following frame P1 | | | | | | | Following frame P2 | | | | | |
|---|---|---|---|---|---|---|---|---|---|---|---|---|---|
| | | CP | AP | AM | CM | Row total | | | CP | AP | AM | CM | Row total |
| Starting frame | CP | | 8 | | | 8 | Starting Frame | CP | | 10 | 10 | 2 | 22 |
| | AP | | 3 | 1 | 2 | 6 | | AP | 2 | | 4 | 2 | 8 |
| | AM | | | | | | | AM | 1 | 1 | | 2 | 4 |
| | CM | | 1 | 1 | | 2 | | CM | 1 | 2 | 2 | | 5 |
| | Column total | | 12 | 2 | 2 | | | Column total | 4 | 13 | 16 | 6 | |

Table 4. Frequency of framing patterns observed in the 2021 cohort. CP = Conceptual Physics; AP = Algorithmic Physics; AM = Algorithmic Mathematics; CM = Conceptual Mathematics. The left section reports data from groups participating in Exercise 1, while the right section refers to those involved in Exercises 2 and 3. The table also includes the qualitative assessment of students' performance in problem solving, classified as successful, neutral, or unsuccessful.

| **Exercise 1** | |
|---|---|
| **P1** | **P2** |

| **Single Students** | **Assessment** | **Groups** | **Assessment** | **Single Students** | **Assessment** | **Groups** | **Assessment** |
|---|---|---|---|---|---|---|---|
| CM-CP | Neutral | CP-CM-AP | Successful | CM-CP-AM-CM-AP-(AM-CM-AP) | Successful | CP-CM-AM-AP | Successful |
| CP-AP-CM | Neutral | AP-CM-AM | Neutral | AM-CM-AP-AM(-CM-AP) | Successful | AP-AM-CM | Neutral |
| CM-AP-CM | Neutral | CP-AP-CM | Successful | CP-CM-AP-AM-(CM-CP) | Successful | CP-AP-CM-AM-CM-CP | Successful |
| AP | Unsuccessful | | | CM-AP-AM | Neutral | | |

| **Exercise 2** | | | | **Exercise 3** | |
|---|---|---|---|---|---|
| **P1** | | **P2** | | **P2** | |
| **Single Students** | **Assessment** | **Single students** | **Assessment** | **Single students** | **Assessment** |
| CP-AP | Neutral | AP-CM | Neutral | AP-CP-AM | Neutral |
| AP | Unsuccessful | AP-AM | Neutral | AP-AM | Neutral |
| CP-AP-AM-CM | Successful | AP-CM-AM-CM | Successful | AP-AM-CM-CP | Successful |
| CP-AP-AM-CM | Successful | AP-CP-AM-CM | Successful | AP-AM-CM-CP | Successful |
| AP-CM-CP-AM | Successful | AP-CM-AM | Successful | AP-AM-CM-CP | Successful |
| CM-CP-AP | Neutral | AP-AM | Neutral | AP-AM | Neutral |
| AP | Unsuccessful | AP-AM | Neutral | CP-AP-AM-CM | Successful |
| CP-AP | Neutral | AP-AM-CP | Successful | AP-AM | Neutral |
| CP-AP | Neutral | CP-AP-AM | Successful | CP-AP-AM-CM | Successful |

Table 5. Frequency of framing patterns in the 2022 cohort. CP = Conceptual Physics; AP = Algorithmic Physics; AM = Algorithmic Mathematics; CM = Conceptual Mathematics. The left section reports data from students participating in Exercise 1, while the right section refers to those involved in Exercises 2 and 3. The table also includes the qualitative assessment of students' performance in problem solving, classified as successful, neutral, or unsuccessful.

| **Exercise 1** | | | |
|---|---|---|---|
| **P1** | **Assessment** | **P2** | **Assessment** |
| AP-CM | Neutral | CM-AP | Neutral |
| AP | Unsuccessful | AP-CM-AM-CM | Successful |
| AP-AM-CM | Neutral | CP-AP-AM | Neutral |
| CP-AP | Neutral | CM-AM | Neutral |
| CM-AM | Neutral | CP-AM-AP | Neutral |
| CP-AP | Neutral | AP-AM | Neutral |
| AP | Unsuccessful | CM-CP-AP-AM | Successful |
| CM-AP | Neutral | | |
| AP | Unsuccessful | | |

| **Exercise 2** | **Exercise 3** |
|---|---|

| P1 | Assessment | P2 | Assessment | P2 | Assessment |
|---|---|---|---|---|---|
| CP-AP | Neutral | CP-AP-AM-CM | Successful | CP-AP-AM | Neutral |
| CP-AP | Neutral | CP-AP-AM-CM | Successful | CP-AP-AM | Neutral |
| CP-AP | Neutral | AM-CM | Neutral | CP-AP-AM | Neutral |
| CP-AP | Neutral | AM-CM-CP-AP | Successful | CP-AP-AM | Neutral |
| CP-AP | Neutral | AP-AM-CP | Neutral | CP-AP-AM | Neutral |
| CP-AP | Neutral | AP-AM-CP | Neutral | CP-AP-AM | Neutral |

The qualitative analysis of students' solving strategies revealed several noteworthy results. The most successful strategies involved comprehensive discussions of the physical scenarios, offering multiple perspectives and solutions that integrated both physical and mathematical reasoning. Pictorial representations were used not only as summaries but also as explanatory tools to visualize the physical context of the mathematical results. In these cases, all six categories of Heller et al. [55] were satisfied, and the five-step problem-solving procedure was consistently applied.

By contrast, less successful approaches were incomplete: diagrams were typically introduced only at the conclusion of the analysis, limiting their effectiveness in enhancing understanding. Neutral strategies tended to concentrate on only one dimension of the problem. Some focused almost exclusively on computational and mathematical aspects, neglecting physical meaning and conceptualization. Others emphasized qualitative reasoning or algorithmic strategies alone, without analyzing the representations or the links between mathematics and physics. This lack of integration between mathematical reasoning and physical visualization was most evident in unsuccessful strategies, where conceptualization and computation were minimal or absent. As a result, the depth of the discussions was significantly restricted.

Importantly, the qualitative analysis of solving strategies was mirrored in the progression of epistemic framing. In successful cases, we observed a fluent transition across the frames described in [31], moving from Conceptual Physics (CP) to Algorithmic Physics (AP), and then to a mathematical treatment combining Algorithmic Math (AM) and Conceptual Math (CM) (see Tables 4 and 5). This coherent cycle was largely absent in neutral and unsuccessful strategies, where framing was fragmented or reversed. These patterns also align with what Kapodistrias et al. (2025) describe as *purposeful transformation* in disciplinary learning [34]: in successful strategies, frame shifts exhibit a progressive narrowing of interpretive possibilities, a foregrounding of the relevant physical–mathematical relations, and a purposeful orientation toward disciplinary goals. In this sense, effective framing dynamics can be read as purposeful reorganizations of reasoning that bring students' approaches closer to expert practice. In particular, a successful solution never uses less than three frames, suggesting that a more frequent use of different frames would lead to better solutions. This further shows that greater flexibility in shifting among frames supports better outcomes. A more detailed investigation – for instance through video recordings of student interactions or short interviews – could provide deeper insights into these dynamics, as also recommended in [31]. Such work will be pursued in future studies.

A key observation in analyzing students' artefacts concerned the way calculations and physical scenarios were described. In the more successful approaches, calculations were clearly articulated through well-developed, descriptive language. These solutions were supported by explicit assumptions and detailed explanations of the physical principles involved, with particular emphasis on complex concepts such as geodesic deviation. This indicates that students not only mastered the steps of mathematical manipulation but also understood the physical significance of these operations. The use of natural, qualitative written language in explaining mathematical procedures played a pivotal role in fostering deeper conceptual understanding [4,41–46]. For example, when employing the CM frame, students drew on concepts from differential geometry to analyze the physical situation. In these cases, their visual representations enabled a shift in perspective from purely geometric reasoning—based on properties of triangles on a sphere or cone—to the underlying dynamics of the process itself.

In contrast, less successful problem-solving strategies were marked by misunderstandings of fundamental physical concepts, such as geodesic motion. For instance, some students referred to the dynamics by stating, "*In General Relativity, bodies follow a geodesic*," which oversimplifies the concept. More accurately, the motion of a body can be described by a geodesic, with the geodesic representing the path that minimizes proper time for an observer. Another example involved a common misconception in optics, often encountered even at the graduate level in physics [62]: "*The trajectory of light rays will be curved by the gravitational field (geodesic motion).*" Here, the student incorrectly described light as rays and framed geodesic motion as the cause of the curvature of light, rather than recognizing it as a consequence of the dynamics.

In the context of cooperative conceptual problem solving, our findings indicate that groups outperformed individual students, aligning with established research on the effectiveness of cooperative learning in higher education [44-46]. Specifically, the three groups we examined employed a comprehensive and well-organized problem-solving strategy, making careful assumptions to explicitly calculate the solution. The use of natural written language was crucial in clearly articulating the steps of the calculation, allowing for a more intuitive and accessible manipulation of mathematical concepts. We noted that, in the context of problem framing, this reasoning was directly linked to the interaction between the CP and CM frames, where the physical context of the problem was translated into mathematical expressions. Ultimately, the solution was presented from a geometric perspective, illustrating the connection between the physical principles and their mathematical representation. An example of successful language use in discussing the physics of the result can be found in a group solution to the P2 problem in the first exercise:

*"We have demonstrated that the shortest path between two points on a sphere is an arc of a great circle, whose radius corresponds to the radius of the sphere (Earth). The minimum distance path can be determined by 'rotating' the Earth so that the new equator aligns with the line connecting the starting and ending points, leveraging the symmetry of the system to show that geodesics on the sphere are great circles. From a physical perspective, we interpret this result by observing that the force exerted on an object moving along the Earth's surface will have only a radial component directed toward the center of the sphere (centripetal force) when following a great circle. For any other path, a tangential component of force would also exist, which serves to slow the motion. Moreover, if we had considered a flat metric, where the manifold has zero curvature, the minimum distance path would have been a straight line connecting the two points."*

In this example, the group addressed the problem by modeling Earth as a sphere and analyzing geodesic motion within this framework to solve for motion along its surface (see Appendix B, Exercise 1, Problem 2). By demonstrating that geodesics on a sphere are great circles, the student effectively transformed the problem from a physical question to a geometric one. The student skillfully integrated geometry and physics, deriving the solution by rotating the Earth and aligning the equator along the trajectory of motion, considering an equatorial arc as the geodesic. This method facilitated the seamless application of classical physics principles, where the forces involved in the motion are connected to the sphere's curvature. Furthermore, from the perspective of GR in both the CM (and subsequently AM) frame, the geodesic's mathematical results are examined from the viewpoint of an observer moving along the geodesic. This illustrates that geodesic motion is a direct consequence of spacetime geometry, linking physical forces with the geometric properties of the sphere. The approach underscores the importance of integrating both mathematical and physical reasoning to fully comprehend the behavior of objects in geodesic motion.

In contrast, less successful problem-solving approaches exhibited notable differences in handling the same material. These approaches often lacked sufficient introduction to equations and failed to explain key concepts or connect them to the physical principles they represent. For instance, although geodesic deviation was mentioned, it was not adequately explained, leaving the physical context ambiguous. In these cases, solutions were framed predominantly in mathematical terms, with a swift transition from physics to mathematics, neglecting clear explanations of the underlying physical concepts. This approach restricted the exploration of the problem, focusing heavily on mathematical formulas while leaving little room for conceptual discussions related to the physical principles. While equations were presented, the process of solving them was not accompanied by a strategic explanation, and solutions were offered purely in mathematical terms, without bridging the gap between mathematics and physics.

Table 6. Frequency of principles and concepts used across the entire sample. For each entry, data from 2021 students are listed first, followed by 2022 students (separated by a comma). Table A (top) refers to Problem-Solving Activity 1 (P1); Table B (bottom) refers to Problem-Solving Activity 2 (P2). Some cells appear empty due to formatting: certain exercises have more entries than others, resulting in blank cells under exercises where no corresponding data are available. A slash (/) indicates that the principle or concept was missing or not present in the student's work.

| A | **Exercise 1** | | **Exercise 2** | |
|---|---|---|---|---|
| | *Principle and concepts* | *Frequency* | *Principle and concepts* | *Frequency* |
| P1 | Locally inertial frame of references | 2 | Non-Euclidean geometry | 7, 3 |
| | Curved spacetime | 4 | Euclidean geometry | 5 |
| | Locally flat spacetime | 4, 2 | Geodesic motion | 9, 5 |
| | Geodesic | | Local observations | 9, 4 |
| | Parallel transport of a vector | 5, 9 | Non-local observations | 8, 4 |
| | locally curved spacetime | 2 | inertial and non-inertial frames of reference | 4, 5 |
| | | | equivalence principle | 2, 5 |
| | | | General covariance principle | 1 |

| B | **Exercise 1** | | **Exercise 2** | | **Exercise 3** | |
|---|---|---|---|---|---|---|
| | *Principle and concepts* | *Frequency* | *Principle and concepts* | *Frequency* | *Principle and concepts* | *Frequency* |
| P2 | Curved spacetime metric | 4, 3 | Newton's gravitational law | 7, 2 | Gravitational redshift | 9, 6 |
| | Geodesic | 6, 6 | Geodesic deviation | 8, 6 | Doppler effect | 10, 6 |
| | curve of minimum distance on a manifold | 2, 2 | geodesic motion | 6, 4 | Schwarzschild metric | 6 |
| | metric | 2 | Riemann tensor | 2 | Equivalence principle | 2 |
| | centripetal force | 1 | gravitational field lines | 1 | Weak field limit | 3, 6 |
| | local inertial frame of reference | 1 | Manifold | /, 3 | | |
| | spherical symmetry | 1 | | | | |
| | Geometrical approximation | /, 1 | | | | |
| | Affine connection | /, 1 | | | | |
| | Manifold | /, 2 | | | | |
| | Maximum circles | /, 1 | | | | |

The order in which principles and content were introduced significantly impacted the approach used to solve the problem and frame the solution (see Tables 4-6). Poor framing often led to incomplete or partially solved problems. Once the physical situation was recognized, students in these approaches frequently resorted to applying standard textbook formulas without providing justification, undermining the depth of their understanding and limiting the quality of their solutions. This observation is supported by the analysis of shifting factors, where poor framing led to a reliance on algorithmic mathematical solving strategies, while richer, physics-based approaches emerged when framing was well-structured (see Table 7). Exploring these shifting frames can guide educators in creating more effective learning environments.

Table 7. Shifting factors. Data corresponds to the entire tutoring activity (Exercises 1 to 3) in 2021 and in 2022 (bold). Empty cells indicate frames that were absent from the analysis.

| | CP | AP | AM | CM |
|---|---|---|---|---|
| CP | | pictorial representations (drawings, diagrams); real physical motion to make assumptions; different physical situation and equations according to descriptions to measure things in different frames<br><br>**pictorial representation (drowning); meaning of geodesic; geometric reasoning; different physical situation according to descriptions to measure things in different frames** | implementing principles and concepts to make an explicit calculation;<br>pictorial representations<br><br>**pictorial representations** | reasoning about geometric properties of spacetime |
| AP | | arguing on specific physical assumptions;<br><br>**different physical situation and equations according to descriptions to measure things in different frames** | reasoning about the maximum and minimum of a function connected to physical situation; different physical situation and equations according to descriptions to measure things in different frames<br><br>**real physical motion to make assumptions; relating physics to mathematics to make explicit calculations** | relying mathematical meaning of what it should be calculated to a specific physical situation<br><br>conceptual aspects of differential geometry to analyze the physical situation<br><br>**assumptions; pictorial and visual representations; curve of minimum distance on a manifold**<br><br>**geodesic motion in two different spacetimes** |
| AM | | Physical assumptions, comparison between travelling on a meridian or parallel<br><br>**Arguing on the sense of results according to physics to exclude solutions and comment the right ones** | | concept of metric (from a mathematical point of view) |
| CM | **Pictorial representation**<br><br>**physics principles** | pictorial representation (drowning)<br><br>**properties of manifolds; relate geometry to physics; pictorial representation of a physics situation** | implementing calculation for geodesic on a sphere | |

During the discussion phase, students were asked how the methodology supported their ability to think and act like experts. Although many reported having encountered problem-solving strategies in their previous studies, most emphasized that this was the first time they had experienced such a structured and conceptually rich approach in a physics course. They highlighted that the proposed methodology differed from their earlier experiences because it was less algorithmic and encouraged them to reflect on the physics before moving to formal calculations. Several students commented that the activity allowed them to "*see physics as never before, deeply engaging with conceptual* *understanding*". While focusing on conceptual aspects rather than immediately applying mathematical procedures felt unfamiliar at first, with practice it helped them to adopt a more expert-like stance. They reported delving more deeply into the underlying physics of the proposed problems and prioritizing physical reasoning, "*as never before*".

Students also shared that this approach prompted them to reconsider their general strategy for studying physics. Whereas they had previously relied primarily on algorithmic procedures, the use of multiple representations encouraged by the activity helped them develop a more comprehensive understanding of the physical scenarios. This shift in perspective was also reflected in our analysis of frame transitions. In particular, students frequently employed pictorial and visual representations during transitions between frames. The visual domain played a key role in bridging conceptual and algorithmic reasoning (see Table 7).

Furthermore, the integration of text-rich problems and accompanying diagrams enabled students to better identify and interpret real physical situations. This encouraged them to consider the physics first, before selecting an appropriate model to describe the phenomena. This change in perspective was clearly evident in their problem-solving strategies: in the more successful approaches, students began by contextualizing and describing the physical scenario—often visually (see Figures 3 and 4)—and only then proceeded to translate it into mathematical formalism.

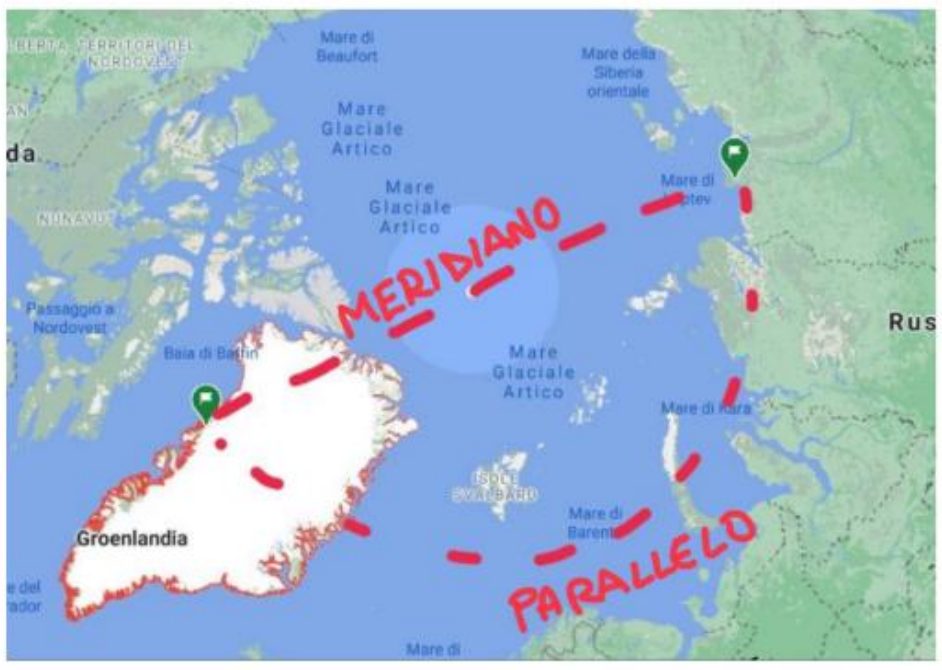

Figura 4.1: Schema dei percorsi Russia-Groenlandia considerati.

Fig. 3 Example of a student's pictorial representation of a travel itinerary in Exercise 1, Problem 2 (P2). The student used "meridians" and "parallels" to illustrate geodesic motion on a spherical surface.

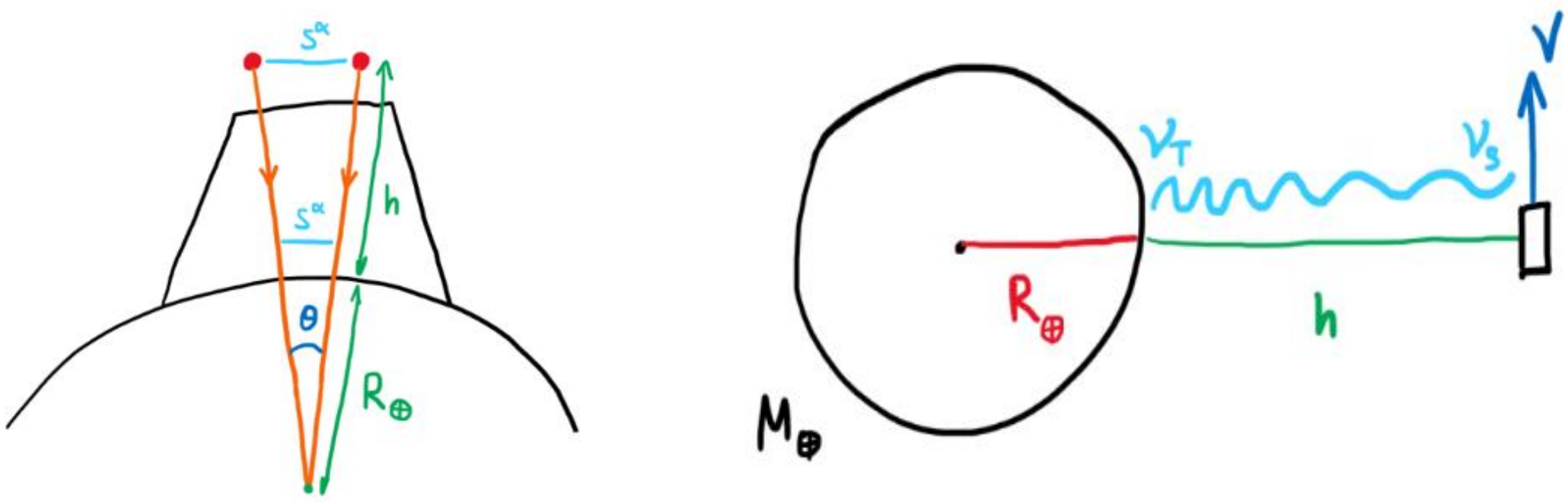

Fig. 4 Student' pictorial representation of geodesic deviation in Exercise 2, P1 (on the left) and of light emission of an atom in curved spacetime as discussed in Exercise 3 (on the right).

In less successful approaches, this initial step was often skipped, and students framed their problem-solving strategies directly within the AM frame. Even when a detailed mathematical analysis was carried out—indicative of engagement with the CM frame—the underlying physical context often remained underdeveloped. As students themselves noted, the use of contextualized problems encouraged them to engage with practical scenarios, prompting reflection on which physical principles were relevant, why they applied, and which specific concepts were required to reach a solution. Engaging with these questions helped establish a strong conceptual foundation—crucial for mastering complex topics such as GR.

It is worth emphasizing that successful strategies were observed predominantly in the 2021 cohort. While we do not have statistical or quantitative data to draw firm conclusions, the lecturer's experience across both cohorts may offer insight into this difference. Notably, in 2021, students worked cooperatively on the problems, whereas in 2022, students worked individually. Furthermore, in-class dynamics revealed a marked contrast in group cohesion: students in 2022 engaged in fewer classroom interactions and showed weaker collaborative behavior compared to the 2021 cohort. These observations suggest that students' learning may benefit significantly from cooperative problem-solving activities—an insight supported by previous studies in the field [31,39,44–46].

The opportunity for students to engage with the tutor online in synchronous, face-to-face training activities and asynchronous formats (comments on the drive with online asynchronous interaction with the tutor) was evaluated as highly beneficial in fostering a deep understanding of the problem. By reviewing comments on files uploaded to the shared folder, students sought clarification of conceptual aspects, such as the direction of emitted light with respect to the observer's frame of reference, to account for deviations from standard perspectives. First-year students found the tutor's role in fostering personalized learning to be crucial, while second-year students did not interact as much, and this lack of engagement was reflected in the overall results, which were largely confined to AM and AP frames. This highlights the importance of tutoring in enhancing personalized learning, as also demonstrated by recent developments in the implementation of Large Language Models in education [63-65]. In summary, students reported that the implementation of this methodology had dual benefits. It provided an opportunity for reflection on their problem-solving strategies, and by engaging with problems conceptually, students were able to identify gaps in their understanding and refine their approaches over time. Writing out their strategies also improved their communication skills, as they were required to articulate their thought processes clearly and coherently, facilitating both their understanding and ability to explain complex concepts.

Our results suggest that the transition from qualitative reasoning to quantitative analysis is not simply a reinforcement of mathematical skills, but also fosters two critical cognitive abilities [23-25]. First, students learn to identify shortcuts once they grasp key principles and concepts. Second, they develop the ability to bridge the gap between intuitive, qualitative reasoning and formal, analytical methods. Importantly, we recommend that this transition be gradual to ensure students fully understand the strategy. A structured approach, including an initial training lecture to explain the model, followed by group discussions and sample solutions, can help students become more comfortable with the process. The primary focus should be on how students engage with equations during the quantitative phase and incorporate them into their problem-solving strategies. Ultimately, this method aims not only to improve students' problem-solving ability but also to cultivate a deeper understanding of the underlying physics concepts, fostering more critical and coherent thinking when applying their knowledge.

Evidence from students' artefacts indicates that, following the first session, most students appropriated the practice of beginning with a diagram and a qualitative rationale, annotated at least some frame shifts in their templates, and coordinated natural-language, pictorial, and symbolic representations more deliberately than before. We also observed an increased tendency to plan transitions from conceptual to algorithmic work and to include evaluation checks (units, limits) prior to finalizing a solution. These developments are consistent with the intended outcomes stated in Section 3 and with the activity's aim to make framing moves visible and tractable in students' written work.

Our investigation into students' problem framing within current models revealed several limitations, highlighting the need for further development in this area. One major challenge is the lack of a structured framework for exploring frames that account more explicitly for the visual and semiotic dimensions of physics learning. Difficulties emerged when attempting to adapt the framework proposed in [31], originally designed to analyze upper-level students' framing in electromagnetism and later slightly revised for studies in quantum mechanics [33]. The adaptation proved challenging due to the framework's limited emphasis on symbolic and visual elements—components that are essential for understanding and solving physics problems effectively.

Another important consideration is that students' problem framing appears to be more cyclical than linear. We noted that students frequently shift back and forth between different frames as they refine their understanding of a problem. To better reflect this dynamic process, a more flexible descriptive account of frame use is warranted. This pattern suggests that learning proceeds through situational, nested cycles, better represented as a spiral than as a straight trajectory [31,47-49]. Developing a richer characterization of these cycles is part of our ongoing work.

We also observed that in less effective approaches, mathematical formalism was often introduced without adequate physical justification, resulting in gaps in conceptual understanding. This underscores the need to more explicitly consider the cyclical nature of problem framing in future models—an issue currently being explored by one of the authors.

Overall, these limitations suggest that a more integrated approach, one that emphasizes both symbolic and visual domains, could enhance students' problem-solving strategies and lead to a deeper understanding of the physical systems under investigation. In particular, natural language—when effectively combined with conceptual reasoning and mathematical formalism—plays a crucial role in developing a richer understanding of physics [66–68].

During the training activity, the trainer emphasized the use of pictorial and diagrammatic representations as a central tool for analyzing the physics underlying the proposed problems. In the first meeting, special attention was paid to identifying which physical quantities were involved in the dynamics and in describing the state of the system through visual representations. This approach, developed collaboratively within the research group, was presented to students as part of the problem-framing methodology introduced in the first training session. However, this was only a qualitative introduction to the integration of semiotic domains in learning. In fact, constructing conceptual understanding requires forming new connections between elements of knowledge, and effective learning demands both behavioral and cognitive engagement. Physics involves multiple semiotic registers—natural language, algebraic expressions, vector diagrams, and pictorial representations—and meaningful learning requires the coordination of these registers [69]. Among them, natural language plays a particularly important role in navigating transitions between representations, such as from algebraic to vectorial formats [69,70]. The limited integration of these domains in existing models suggests that placing greater emphasis on the explanation of physical principles and their mathematical representations could significantly improve students' conceptual understanding and problem-solving performance.

## Conclusion

In this paper, we examined the effectiveness of conceptual problem solving in supporting the learning of complex topics such as general relativity (GR). Drawing on previous research in the field, we presented a case study that investigated students' problem framing during a training activity conducted at the Department of Physics, University of Cagliari (Italy), in 2021 and 2022. Our research focused on how students structure their problem framing in conceptual problem-solving, particularly in GR (RQ1). We explored how this framing contributes to their conceptual learning and understanding of GR (RQ2). Additionally, we assessed whether our methodology effectively evaluates students' problem-solving and framing competencies, providing valuable insights into their learning process (RQ3).

Our results showed that an effective approach to teaching requires more than simply applying mathematical techniques—it demands a careful integration of conceptual understanding with mathematical formalism. Through the development of a pedagogical framework that emphasizes conceptual problem-solving and the blending of qualitative reasoning with quantitative techniques, we have shown that students are better equipped to grasp the underlying physical principles and apply them to solve problems in GR. The results of our investigation confirm that students who are able to frame problems conceptually, justify their choices based on physical principles, and move fluidly between intuitive reasoning and formal mathematical procedures exhibit a deeper understanding of the material. Our findings also corroborate the interpretation of students' transitions between epistemic frames as *purposeful transformations* – goal-oriented reorganizations of thought that progressively narrow interpretive possibilities, highlighting relevant relationships in reasoning, and align problem-solving processes with the epistemic aims of physics [34].

By focusing on epistemological framing, students are encouraged to see mathematics not just as a set of formulas to be applied, but as a language that connects abstract concepts to real-world phenomena. This approach fosters the development of symbol sense, which is a key skill for students to reason effectively about physical systems and navigate complex problems (RQ1).

Our research addresses the central question of how conceptual problem framing influences learning in GR (RQ2). The findings show that by encouraging students to develop a robust conceptual framework, we can help them build a more coherent and comprehensive understanding of the physical world, allowing them to approach problems with greater flexibility and insight. Moreover, the ability to seamlessly integrate conceptual and mathematical reasoning empowers students to not only solve problems but to communicate their solutions clearly and effectively—a skill that is invaluable in both academic and professional settings.

Additionally, our findings suggest that our methodology provides instructors with valuable insights into students' thought processes (RQ3). In particular, it offers an opportunity to identify misconceptions early and provide more targeted, constructive feedback. By focusing on the process of problem framing rather than simply the final answer, instructors can better understand the areas where students may be struggling and help them overcome these obstacles more effectively. This allows for more targeted feedback and guidance, which can be much more effective than just looking at a final answer. Ultimately, the goal is to help students see the bigger picture—understanding the concepts deeply, not just solving equations mechanically. This is corroborated by both the analysis of students' artefacts and their feedback on the training activity, which suggest that they began to internalize (i) diagram-first problem analysis, (ii) explicit qualitative justification, (iii) planned transitions to algorithmic steps, and (iv) basic checks on mathematical form. Overall, our findings indicate that integrating problem framing into conceptual problem solving provides a coherent means both to investigate students' problem-solving processes and to enhance their skills. Nevertheless, future work will refine these practices and examine their stability across tasks and settings.

However, our investigation revealed challenges in the current models of problem framing, particularly due to the lack of a structured framework that integrates visual and symbolic elements crucial for understanding physics [65-67]. Moreover, students' problem framing is better understood as a cyclical process, with students revisiting previous steps as they refine their understanding. Less effective approaches often introduce mathematics without physical justification, leading to gaps in conceptual understanding. We suggest that a more integrated approach, emphasizing symbolic and visual elements alongside natural language, would improve problem-solving and foster deeper conceptual understanding in physics. This is left for future investigations.

In conclusion, we assert that the key to mastering complex physics topics, such as GR, lies in developing students' ability to blend qualitative and quantitative reasoning [23-25]. The pedagogical framework presented here encourages students to engage with both the mathematical formalism and the underlying physical principles, fostering a deeper and more integrated understanding of the subject and pedagogical matter. By cultivating this balance, we prepare students not only to solve specific problems but to think critically and creatively as physicists, bridging the gap between abstract theory and real-world applications. This holistic approach to problem-solving ultimately equips students with the cognitive tools necessary to navigate the challenges of advanced physics and think like experts.

**Acknowledgments**

We are grateful to all the students who participated in the study. We also thank the anonymous referees for their constructive feedback, which helped us improve the quality of the manuscript.

**Ethical statement**

Informed consent to participate in the study has been obtained from participants. Any identifiable individuals participating at the study have been also aware of intended publication. Informed consent to publish has been obtained from participants of the study. This work was carried out in accordance with the principles outlined in the journal's ethical policy and with the "Codice etico e di comportamento" of the University of Cagliari.

**Funding declarations and conflicts of interests**

There are no known conflicts of interest associated with this publication and there has been no significant financial support for this work that could have influenced its outcome.

**Declaration of generative AI in scientific writing**

The authors declare that no generative artificial intelligence tools were used in the design, analysis, or interpretation of this study. Language editing assistance was limited to grammar and style correction using AI-assisted tools (ChatGPT-5), with all intellectual content, analysis, and interpretation solely the responsibility of the authors.

## Appendix A

**Exercise**

**1. Identify the principles and concepts necessary to solve the problem.**

Text of the problem

**Choose from the following options (multiple choice)**

| Principles | Concepts |
|---|---|
| a) … | a) … |
| b) … | b) … |
| c) … | c) … |
| d) … | d) … |
| e) … | e) … |

**Justify your choice (max 500 words, no formulae)**

**2. Conceptual problem and analytic solution**

Text of the problem

***Conceptual scheme***
*Principle and concepts (make a list, no formulae)*

*Justify your choice (max 500 words, no formulae)*

***Analytic Scheme***
- Draw the physical situation expressed in the problem.
- Plan a solution strategy.
- Execute the solution strategy.
- Discuss the result in light of the concepts and principles identified.

## Appendix B

**Exercise 0**

Problem: You're at the bar with a colleague, and you start discussing some intriguing aspects of Einstein's special relativity. After a few too many drinks, you both decide to do a thought experiment. You imagine building a tower, with height H, next to the Physics Department at the University of Cagliari. Your job is to climb to the top and drop a particle of mass m. Your colleague stays on the ground and decides to measure the energy of the particle when it hits the ground. In this thought experiment, they magically convert all the energy into a single photon, with no energy loss (let's not worry about the details). The photon is then sent back up to the top of the tower, to the same height H. When it arrives, you measure its energy and find it to be E'. Miraculously, you can also convert the photon back into a particle of mass m' = E', and to keep the experiment from turning into the discovery of perpetual motion, you confirm that m' = m. Now, you decide to compare the energy of the photon measured on the ground, E, with the energy E' measured when it reaches the top. You realize that the frequency of the photon has shifted, specifically towards the red end of the spectrum.

You both want to figure out what happened and find an explanation. Excited by the results, you also wonder what would happen if:

1. The experiment was conducted without the Earth's gravitational field;
2. You and your colleague each conducted the experiment independently, still in the presence of the Earth's gravitational field;
3. The experiment was done as in point 2, but both you and your colleague were in free fall within the gravitational field.

Finally, you discuss how important it is to compare measurements made at different points in space and the need to clearly define how to measure a physical quantity to avoid paradoxical conclusions.

| Principles | Concepts |
|---|---|
| a) Energy conservation | a) Wave-particle duality |
| b) First principle of dynamics | b) Gravitational redshift |
| c) Inertial mass – gravitational mass equivalence | c) Light quantization |
| d) Mass energy equivalence | d) Uniformly accelerated motion |
| e) Relativity principle | e) Locality of the measure process |

**Exercise 1**

Problem 1 (P1): One of the revolutionary consequences of Einstein's general relativity can be summed up in the words of physicist John Wheeler: "Matter tells spacetime how to curve, and spacetime tells matter how to move." Driven by a highly pragmatic approach, you decide to experience firsthand the meaning of the phrase "spacetime tells matter how to move." To do this, you embark on a scientific meditation journey that takes you from Valencia to Athens, with a stop in Amsterdam, before returning to Valencia. You carry a backpack with a flag sticking out of the top, pointing upward. You wonder what happens to the direction of the flag during your journey, both if you are the one documenting its position along the way, and if it's an external observer watching you from the Moon. (Assume that you are point-like.)

| Principles | Concepts |
|---|---|
| a) General covariance principle | a) Parallel transport |
| b) Equivalence principle | b) Gravitational redshift |
| | c) inertial frames of reference |
| | d) spacetime locally flat |

Problem 2 (P2): After the long lockdown caused by the pandemic, you want to plan a boat trip that will allow you to enjoy cooler climates even in the summer. You decide to head to higher latitudes and visit the Arctic Circle. Your travel itinerary includes departing from the Russian coast, heading to Greenland, and then returning to your port of departure. Time is limited for your vacation, so you need to decide which route to take in order to move according to your plan and in the shortest time possible. Since you're studying for the general relativity course, you know that a path along the equator would be ideal if you were planning to visit the countries of the equatorial belt, but unfortunately, it's the cooler climates that you prefer. Driven by curiosity, before you depart, you decide to identify (and prove) which trajectory is most suitable for your trip and reflect on the differences with other possible routes in light of the knowledge you've acquired in the course. (For simplicity, focus on just two possible paths with different geometries.)

**Exercise 2**

Problem 1 (P1): Live your dream of becoming a physicist by landing a job at the European Space Agency. One of the agency's programs involves testing a new method for transmitting information inside space shuttles using light signals sent from one point to another across the cabin, where the control panels are located. You've been asked to join the astronauts on board to observe and describe the phenomenon. The emitter and receiver are positioned at opposite ends of the cabin, with the receiver being a fluorescent wall. Your task is to determine the exact position of the light signal on the receiver in three different scenarios: when the shuttle is on the launch pad on Earth's surface, during flight as it moves away from Earth (accelerating at g, where g is the Earth's gravitational acceleration), and in orbit around Earth. For each situation, identify the type of motion the light signal undergoes. Finally, compare your measurements with those of a colleague on the ground. (Assume light behaves as a particle and that g remains constant.)

| Principles | Concepts |
|---|---|
| a) General covariance principle | a) Euclidean geometries |
| b) Equivalence principle | b) Non-euclidean geometries |
| | c) inertial frames of reference |
| | d) non-inertial frames of reference |
| | e) geodesic motion |
| | f) local observations |
| | g) non-local observations |

Problem 2 (P2): Driven by your irrepressible skepticism, you decide to test your knowledge of general relativity with a thought experiment. Specifically, you are interested in the distortion effects on relative distances between bodies in free fall within gravitational fields. You and a friend embark on an expedition to Mount Everest, bringing two heavy objects. One of you climbs to the top of the mountain, while the other stays at the base. From the summit, you drop the two objects simultaneously toward the ground (assuming no natural obstructions to their fall, and that you can indeed reach the top of Everest – this is a thought experiment!). You know that the effects of Earth's gravitational field, though weak, are not negligible. To probe these effects, you decide to focus solely on the evolution of the relative separation between the trajectories of the two objects (without delving into the detailed path of each object) from the mountain summit to the ground. First, you compare the evolution of this relative distance using Newtonian mechanics, and then apply general relativity. Discuss the physical result predicted by both theories, emphasizing the role of the curvature tensor in the trajectories of the falling objects. (In the case of general relativity, identify the relative distance with a time-like vector $s^{\alpha}$).

**Exercise 3**

Problem 2 (P2): The European Space Station has launched a call for astronauts to conduct a crucial experiment on Einstein's General Relativity. The research focuses on studying atomic transitions within a gravitational field. Fueled by your unwavering enthusiasm, you and a colleague decide to apply for the mission. Thanks to your abilities, you are both selected to fulfill the task of being astronauts. The more courageous of you will go to the space station, while the other will stay on Earth to work in the laboratory.

Both of you are asked to observe the same experiment: in the Earth-based laboratory, an atom emits a photon with frequency $\nu_T$ following an atomic transition induced by Compton scattering. The photon reaches the space station with frequency $\nu_S$. The space station is located at a height $h$ above the laboratory and moves with a tangential velocity $v_S$. From the laboratory on Earth, you are asked to communicate the value of the photon frequency received at the space station and compare it with the value measured on Earth. Additionally, from the Earth-based laboratory, you are asked to calculate the frequency shift $\Delta\nu/\nu_T$ in the weak gravitational field and low-speed limit.

Finally, you are informed that you must conduct an experiment for which you were not prepared, but for which you have all the necessary resources—food, energy, technology, and psychological support - to carry it out. They propose that you ignite the engines and head toward a Schwarzschild black hole in our galaxy (remember, it's a thought experiment). The black hole has mass $M$ and radius $R$. You are asked to repeat the experiment while in a stable orbit at distance $h_0$. From this position, you receive a photon emitted with frequency $\nu_{BH}$ from electron-positron annihilation near the event horizon of the black hole. The photon arrives at your receiver with frequency $\nu_{0S}$ (you are in a stable orbit around the black hole and moving with tangential velocity $v_{0S}$ ). You measure the change in frequency $\Delta\nu/\nu_{BH}=(\nu_{0S}-\nu_{BH})/\nu_{BH}$ as a function of the black hole's radius and your distance from it. Comment on your results, as they will be crucial for developing future quantum gravity measuring devices in space. Afterward, you can return to Earth and enjoy the applause.